\documentstyle[twoside,fleqn,npb,epsfig]{article}
\newcommand{\AmS}{{\protect\the\textfont2
  A\kern-.1667em\lower.5ex\hbox{M}\kern-.125emS}}

\title{\phantom{}\vskip -1.2 cm
 \hfill CERN-TH/2002-316
 \vskip 0.01 cm
 \hfill TTP02-35
 \vskip 0.01 cm
 Radiative return at \(e^+e^-\) factories.
 \thanks{Work supported in part by BMBF under grant number 05HT9VKB0, 
EC 5-th Framework EURIDICE network  project HPRN-CT2002-00311 and
 TARI project HPRI-CT-1999-00088;
 presented by H. Czy{\.z} at RADCOR 2002, Kloster Banz, 
 September 8 - 13, 2002.}}

\author{
Henryk Czy\.z\address{{Institute of Physics, University of Silesia,
PL-40007 Katowice, Poland.}}$^{,}$
\thanks{Supported in part by 
EC 5-th Framework, contract HPRN-CT-2000-00149; e-mail:czyz@us.edu.pl}%
Johann H. K\"uhn\address{Institut f\"ur Theoretische Teilchenphysik,
Universit\"at Karlsruhe, D-76128 Karlsruhe, Germany.}$^{,}$
\thanks{e-mail:jk@particle.uni-karlsruhe.de}%
Germ\'an Rodrigo\address{Theory Division,
  CERN, CH-1211 Geneva 23, Switzerland.}$^{,}$
\thanks{Supported in part by 
E.U. TMR grant HPMF-CT-2000-00989; e-mail: german.rodrigo@cern.ch}
  }

\begin{document}

\begin{abstract}
 The energy dependence of the electron - positron hadronic cross section
 can be measured not only by a straightforward energy scan, but also
 by means of the radiative return method. To provide extensive
 comparisons between theory and experiment a Monte Carlo event generator
 is an indispensable tool. We have developed such a generator called
 PHOKHARA, which simulates 
 \(e^+e^-\to {\rm mesons} + {\rm photon(s)}\) processes.
 In this paper we present its latest tests and upgrades.
    
\end{abstract}

\maketitle

\section{INTRODUCTION AND PHOKHARA UPGRADES}

 In view of the precision of
the recent measurements of the muon anomalous magnetic moment
$a_{\mu} \equiv (g-2)_{\mu}/2$ at BNL~\cite{Brown:2001mg},
hadronic contributions are crucial for the interpretation of
this measurement, in particular for the isolation of
the electroweak or of non-Standard Model
physics contributions~\cite{Hughes:1999fp}.
Their understanding becomes even more important, as it seems
\cite{DEHZ} that the \(e^+e^-\) annihilation data are not consistent with
 the \(\tau\) decay data.
A new \(a_{\mu}\) measurement, which is under way, will challenge the
theoretical predictions even more.

 An important ingredient and the dominant source of uncertainties
in the theoretical prediction 
for the muon anomalous magnetic moment 
is the hadronic vacuum polarization ~\cite{hadronicmuon}.
It is in turn related via dispersion relations to the 
cross section for electron--positron annihilation into hadrons
$\sigma_{had}=\sigma(e^+ e^- \rightarrow hadrons)$ and in some cases
 via isospin symmetry to the decay width $\tau \to \nu_{\tau}+hadrons$. 
This quantity plays an important role also in the 
evolution of the electromagnetic coupling $\alpha_{QED}$ from the 
 Thompson limit to high energies~\cite{hadronicmuon,runningQED}.
The interpretation of improved measurements 
at high energy colliders such as LEP, Tevatron, the LHC or TESLA  
 therefore depends significantly on the precise knowledge of $\sigma_{had}$.

The feasibility of using tagged photon events at high luminosity
electron--positron storage rings, such as the $\phi$-factory, DA$\Phi$NE,
CLEO-C or $B$-factories, to measure $\sigma_{had}$ over a wide range of 
energies has been proposed and studied in detail 
in~\cite{Binner:1999bt,Melnikov:2000gs,Czyz:2000wh} 
(see also~\cite{Spagnolo:1999mt,Khoze:2001fs}).
The machine is operating at a fixed energy of the 
 \(e^+e^-\) centre-of-mass system (cms)
 and the initial state radiation (ISR) is used to reduce
 the invariant mass of the hadronic system.  

\begin{figure}[htb]
\vspace{9pt}
\epsfig{file=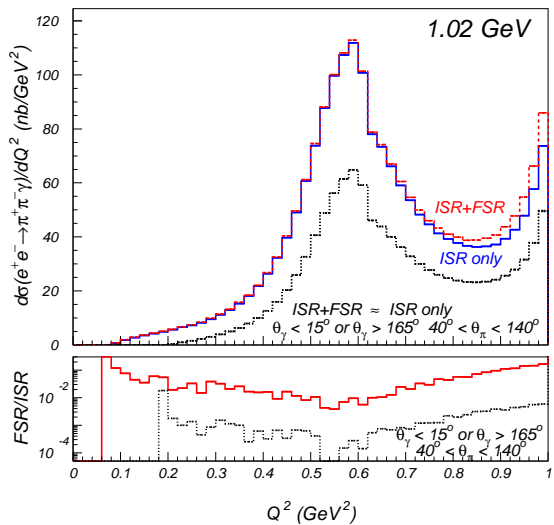,width=8 cm,}\vspace{-4. cm}
\caption{The role of the cuts in the suppression of the FSR contributions
to the cross section. Results from the PHOKHARA generator. 
No cuts (upper curves) and suitable cuts applied (lower curves). }
\label{fig:isrtofsr}
\end{figure}

The radiation of photons 
from the hadronic system (the final state radiation, FSR) 
should be considered as a background and can be suppressed by choosing 
suitable kinematical cuts, 
or controlled by the simulation, once a 
suitable model for this amplitude has been adopted. 
One finds that 
selecting events with the tagged photons close to the beam axis and well 
separated from the hadrons indeed reduces FSR drastically. As
demonstrated in Fig. \ref{fig:isrtofsr},
 the FSR contribution to the total cross section 
can be easily reduced to the 1\% level. 
 The model dependence of the FSR in the case of the \(\pi^+\pi^-\) final state
 can be controlled by the same experiment through studies of the
 forward--backward asymmetry of the angular charged pion distribution.
 The asymmetry comes from FSR--ISR interference 
 and integrates to zero for 'charge blind' configurations. As a result
 it does not contributes to rates in Fig. \ref{fig:isrtofsr}. It can be
 used, however, to calibrate the FSR amplitude and more detailed
 tests of its model dependence.
\vspace{-1. cm}

 When running at higher energies,  
 the FSR is suppressed with respect to the ISR by the different 
 behaviour of various propagators and form factors relevant to the problem.
 In practice it means that no special angular cuts are needed
 to suppress the FSR contribution when running at high energies.

The suppression of the FSR overcomes the problem of its model 
dependence, which must be taken into account in a completely inclusive 
measurement~\cite{Hoefer:2001mx}.

Preliminary experimental results using this method have 
been presented recently by the KLOE collaboration at 
DA$\Phi$NE~\cite{Aloisio:2001xq,Denig:2001ra,Adinolfi:2000fv,Denig:2002}.
Large event rates were also observed by the BaBar 
collaboration~\cite{babar}.

In the first version of the newly developed Monte Carlo program 
PHOKHARA \cite{RCKS}
 we have considered the full next-to-leading order (NLO) QED
corrections to the ISR in the annihilation 
process $e^+ e^- \rightarrow \gamma + hadrons$, for the case 
where the photon is 
observed under a non-vanishing angle relative to the beam direction. 
The virtual and soft photon corrections were presented in \cite{Rodrigo:2001jr}
and the contribution of the emission of a second hard photon
in  \cite{RCKS}. The final hadronic state was limited to the \(\pi^+\pi^-\),
with the hadronic current modeled as in \cite{Kuhn:1990ad}, 
and the final state emission was not included. The program allowed also
for the generation of \(\mu^+\mu^-\gamma(\gamma)\) final states, again
limited to the emission of photon(s) from the initial leptons.

\begin{figure}[htb]
\epsfig{file=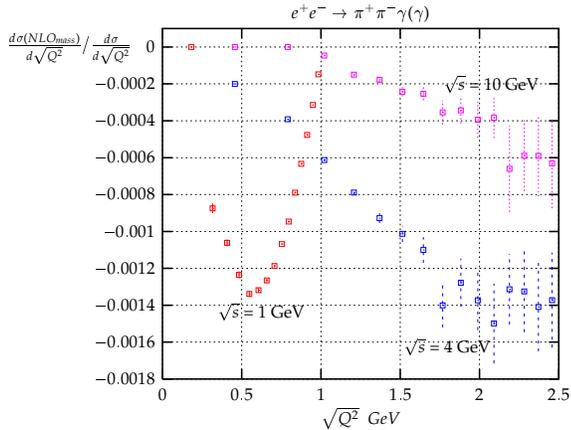,width=7.5 cm,}
\caption{The relative contributions of the non-leading mass corrections to the
 differential cross section at \(\sqrt{s}\) = 1 GeV, 4 GeV and 10 GeV. }
\label{fig:nlo_mass}
\end{figure}

Radiative
corrections proportional to (\(\alpha m_e^2\) ) 
relevant to configurations with photons emitted at very 
low (\(\simeq m_e/\sqrt{s}\))
 angles relative to the beam direction were calculated in \cite{RK02}
and are included in the new version of PHOKHARA 
\cite{CGKR}. The leading order corrections proportional to \(m_e^2\)
are typically of the order of a few per cent \cite{Rodrigo:2001cc}, while
the non-leading ones are of order of 0.1\%, as seen from 
Fig.\ref{fig:nlo_mass}. They will be important when the precision of
 the measurement will be below 1\%. 
Their effect depends on \(Q^2\) and thus affects
  the \(Q^2\) distribution  from which the hadronic
 cross section is extracted.
\begin{figure}[htb]
\epsfig{file=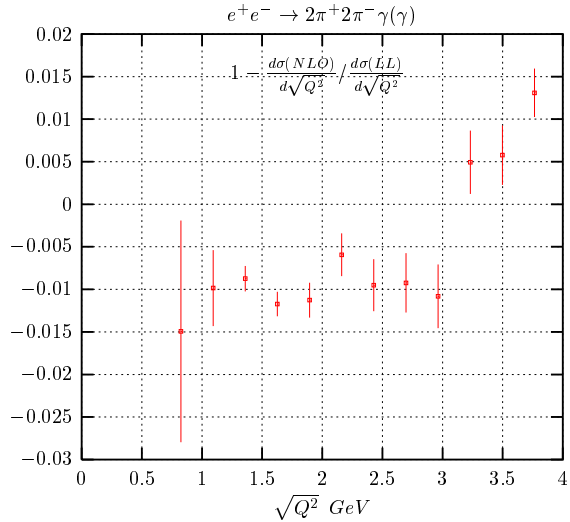,width=7.5 cm,}
\caption{The relative 
non-leading contributions to the differential cross section 
at \(\sqrt{s}\) = 4 GeV. NLO - full next-to-leading result, 
LL - leading logarithmic approximation.  }
\label{fig:ll_nlo_1}
\end{figure}

 Another new feature of the PHOKHARA event generator is the inclusion
 of the four-pion hadronic final states (\(2\pi^+ 2\pi^-\) and 
 \( 2\pi^0\pi^+\pi^-\) ). The description of the hadronic current in that
 case is based on the paper \cite{Fink}, with changes described in 
\cite{Czyz:2000wh}. 
The comparison with the Monte Carlo, which simulates 
the same process at leading order \cite{Czyz:2000wh}
  and includes additional collinear 
radiation through structure function (SF) techniques, shows typical difference
of order of 1\% as seen in Fig. \ref{fig:ll_nlo_1} (a similar behaviour can be
observed for \( 2\pi^0\pi^+\pi^-\) final state).
 The non-leading contributions to the cross section of the four-pion
 final states are of the expected
 size and of the same order as for the two-pion final state \cite{RCKS}.

 The program now includes also the contributions from the final state
 emitted photons together with the ISR--FSR interference calculated at 
 the lowest order for \(\pi^+\pi^-\) and \(\mu^+\mu^-\) final states,
 while for the four-pion final states the FSR contribution is not taken into
 account.

\section{TESTS OF PHOKHARA}

 An obvious and one of the most important tasks
 in the construction of a Monte Carlo event generator is to demonstrate
 that its technical accuracy is much better than the desired physical
 accuracy. The tests that were performed for the previous version of
 PHOKHARA \cite{RCKS} are still valid, within the limitations
 of this version, which was applicable for non-vanishing photon angles. 
\begin{figure}[htb]
\epsfig{file=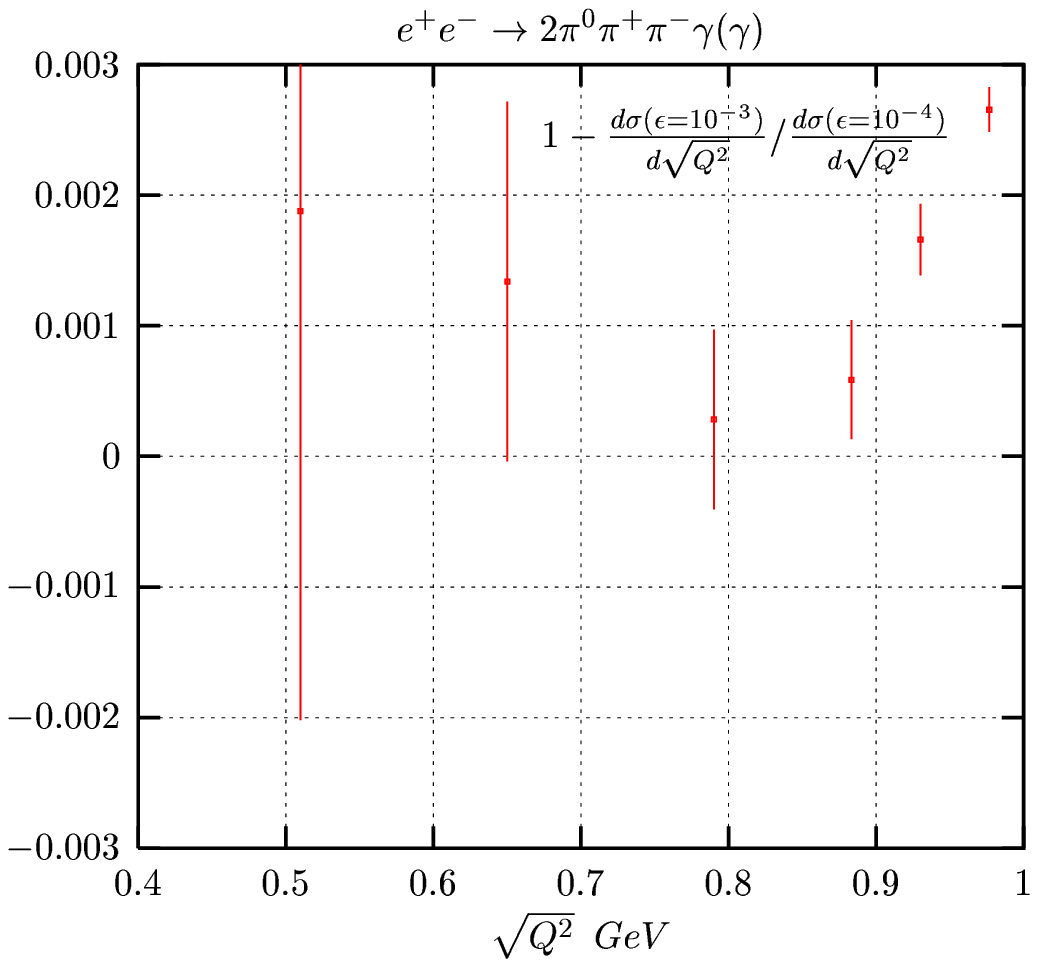,width=7.5 cm,}
\caption{The relative difference of the differential cross sections for two
 different values of the separation parameter \(\epsilon\).}
\label{fig:e43}
\end{figure}
\begin{figure}[htb]
\epsfig{file=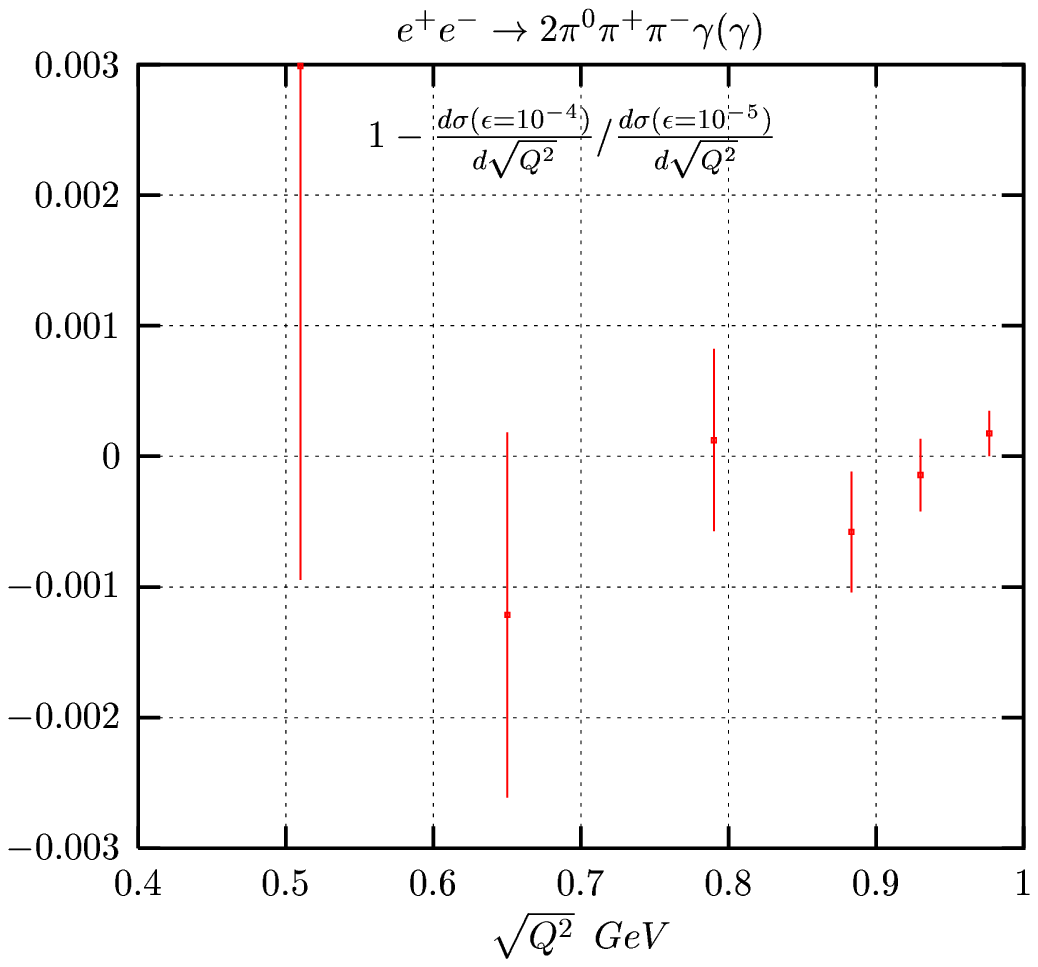,width=7.5 cm,}
\caption{The relative difference of the differential cross sections for two
 different values of the separation parameter \(\epsilon\).}
\label{fig:e54}
\end{figure}

 In the first step we demonstrate
 the independence of total and differential cross sections 
 of the separation parameter \(\epsilon\) (called \(w\) in \cite{RCKS})
  between soft and hard photon
 regions. The soft photon contribution is calculated analytically,
 while the additional hard photon is treated
 via Monte Carlo simulation. The parameter that specifies the
 separation between the
 two regions of the phase space
 has to be kept small enough to validate the soft photon 
 approximation and large enough to avoid negative weights. 
 We performed the tests for
 a \(\pi^+\pi^-\) hadronic final state in \cite{RCKS}, while for one of the
 four-pion modes the results are collected in Figs. \ref{fig:e43} and 
 \ref{fig:e54}. From Fig. \ref{fig:e43} it is clear that the choice
 \(\epsilon=10^{-3}\) is still too big, whereas Fig. \ref{fig:e54} demonstrates
 the stability of the results between \(\epsilon=10^{-4}\)
 and \(\epsilon=10^{-5}\). This also proves that the Monte Carlo integration
 works well in the soft photon region.
\begin{figure}[htb]
\epsfig{file=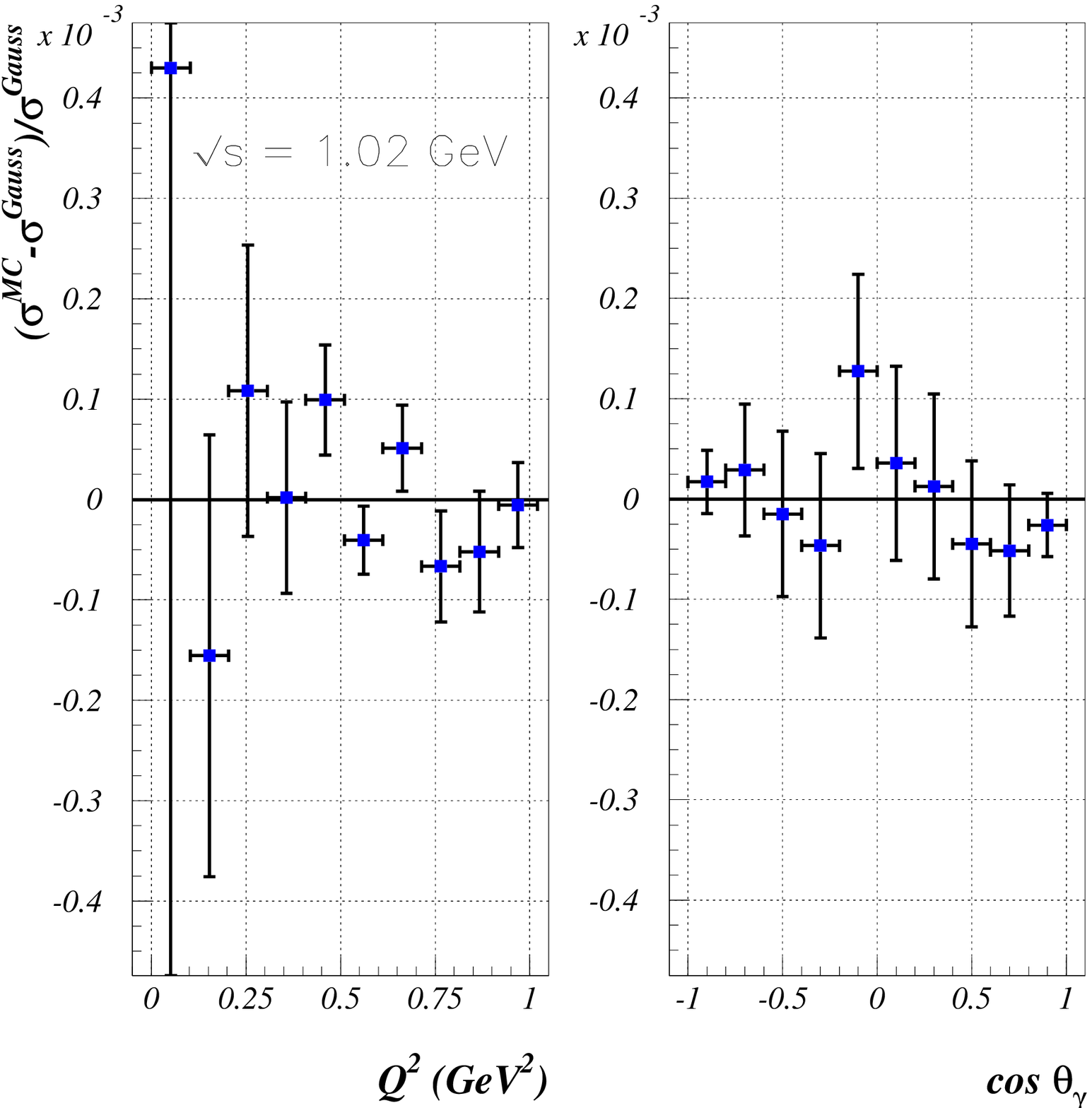,width=7.5 cm,}
\caption{The relative difference between differential cross sections
 obtained by the PHOKHARA Monte Carlo generator (MC) and a Gauss numerical
 integration (Gauss).}
\label{fig:Fig3}
\end{figure}
 The Monte Carlo integration of the part of the program that simulates
 one hard large-angle photon 
 emission was tested in \cite{RCKS} 
 against a Gauss numerical integration. As shown in Fig. \ref{fig:Fig3}
 a technical precision of the program at the level of \(10^{-4}\) was 
 demonstrated.
\begin{figure}[htb]
\epsfig{file=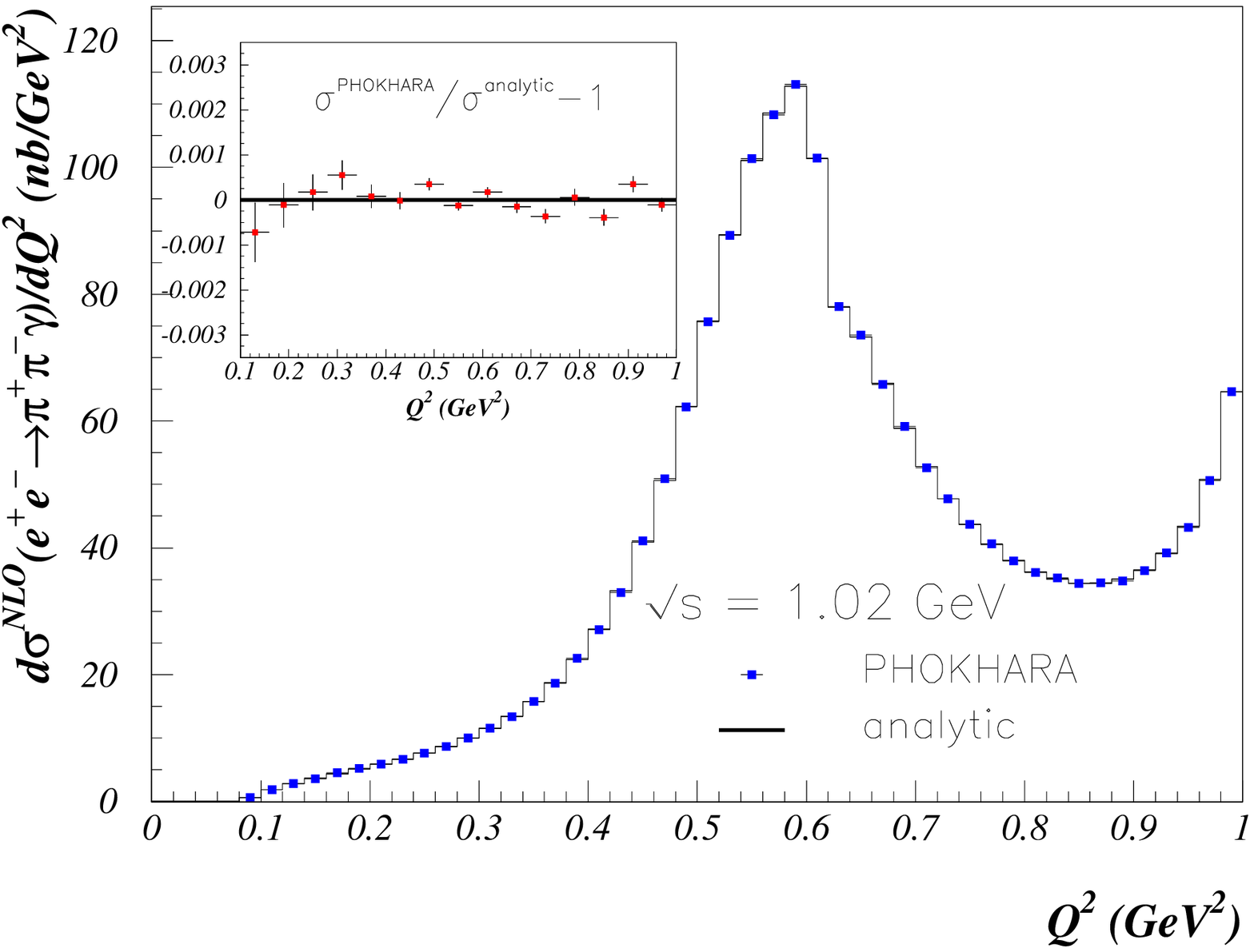,width= 8 cm,}
\caption{A comparison between PHOKHARA and analytical \cite{BNB} results. }
\label{fig:small_1gev}
\end{figure}
 Analytical results exist for the differential (in \(Q^2\)) cross section
 integrated over the whole angular range of the photon(s) for both one and
 two emitted photons \cite{BNB}. The comparison with PHOKHARA
 can be found in Fig. \ref{fig:small_1gev}, demonstrating again an excellent
 technical precision also for the two-photon final state.
 The results presented in Fig. \ref{fig:small_1gev}
 refer to the sum of virtual and hard corrections to the
 \(e^+e^-\to\pi^+\pi^-\gamma\) cross section, while
 more detailed tests can be found in \cite{CGKR}.

\section{CONCLUSIONS}

 The PHOKHARA Monte Carlo event generator was upgraded, allowing
 for simulation in the small photon angles region.
 Besides the \(\pi^+\pi^-\) hadronic final state, 
 its present version includes also
 \(2\pi^+2\pi^-\) and \(2\pi^0\pi^+\pi^-\) final states. 
 For the \(\pi^+\pi^-\) and \(\mu^+\mu^-\) final states, the FSR photonic
 contributions were implemented at the lowest order, 
 including ISR--FSR interference.
 Further upgrades
 are in progress.

\end{document}